\documentclass[12pt]{article}

\usepackage[utf8]{inputenc}
\usepackage[T1]{fontenc}

\usepackage{setspace} 
\usepackage[vmargin = 1.25in, hmargin = 1.25in]{geometry}

\usepackage[usenames,dvipsnames,svgnames]{xcolor}
\usepackage{tensor}
\usepackage{amsmath, amssymb, amsfonts, graphicx, tikz,pdflscape, mathtools, amsthm, bm,multirow, multicol, booktabs}
\usepackage{verbatim}
\usepackage{natbib}
\usepackage{accents}
\newcommand{\ubar}[1]{\underaccent{\bar}{#1}}

\usepackage{comment}
\usepackage[inline]{enumitem}
\usepackage[skip = 10pt]{caption}
\allowdisplaybreaks[1]
\allowdisplaybreaks[1]

\definecolor{Blue}{RGB}{86,180,233}
\definecolor{Orange}{RGB}{230,159,0}
\definecolor{Green}{RGB}{0,158,115}
\definecolor{GmailBlue}{RGB}{42, 93, 176} 
\usepackage[
colorlinks=true,
citecolor= GmailBlue,
linkcolor=GmailBlue,
urlcolor = GmailBlue
]{hyperref}
\newcommand{\bibtexorder}[1]{}

\usepackage{pgfplots}
\usepgfplotslibrary{groupplots}
\usepackage{tikz}
\usetikzlibrary{matrix,calc,shapes,arrows.meta,positioning}
\usetikzlibrary{plotmarks}
\pgfplotsset{width = \textwidth/2}
\tikzstyle{hollow}=[circle,draw,inner sep=1.5]
\tikzstyle{solid}=[circle,draw,inner sep=1.5,fill=black]
\pgfplotsset{compat = newest}

\usepackage[capitalize,noabbrev]{cleveref}
\AddToHook{cmd/appendix/before}{%
  \crefalias{section}{appendix}%
}
\newtheoremstyle{breakital}
{}
{}
{\itshape}
{}
{\bfseries}
{}
{\newline}
{}

\theoremstyle{breakital}
\newtheorem{thm}{Theorem}
\newtheorem*{theorem*}{Theorem}
\newtheorem*{cor*}{Corollary}

\newtheorem{lem}{Lemma}

\crefname{prop}{Proposition}{Propositions}
\crefname{thm}{Theorem}{Theorems}
\crefname{lem}{Lemma}{Lemmas}

\newtheoremstyle{break}
{}
{}
{}
{}
{\bfseries}
{}
{\newline}
{}

\theoremstyle{break}

\crefname{as}{Assumption}{Assumptions}

\theoremstyle{definition}

\newtheorem{rem}{Remark}
\crefname{rem}{Remark}{Remarks}
\newtheorem*{rem*}{Remark}

\numberwithin{lem2}{section}
\crefname{lem2}{Lemma}{Lemmas}

\def\a{\alpha}
\def\b{\beta}

\def\e{\varepsilon}

\def\th{\theta}

\label{key}

\def\D{\Delta}
\def\Th{\Theta}

\def\R{\mathbf{R}}

\def\PP{\mathcal{P}}

\def\XX{\mathcal{X}}

\def\P{\mathbf{P}}

\DeclareMathOperator{\E}{\mathbf{E}}
\DeclareMathOperator{\supp}{supp} 
\DeclareMathOperator*{\argmax}{argmax}

\newcommand{\abs}[1]{\lvert #1 \rvert} 

\newcommand{\Paren}[1]{\left( #1 \right)}

\newcommand{\Brac}[1]{\left[ #1 \right]}

\newcommand{\Set}[1]{\left\{ #1 \right\}}

\usepackage{pgfplots}
\definecolor{amethyst}{rgb}{0.6, 0.4, 0.8}
\definecolor{babypink}{rgb}{0.96, 0.76, 0.76}

\usepgfplotslibrary{fillbetween}

\def\addlegendimage{\csname pgfplots@addlegendimage\endcsname}

\usepackage{subcaption}

\usepackage{mathtools}

\everymath{\displaystyle}

\usepackage{datetime}
\newdateformat{specialdate}{\THEDAY~\monthname[\THEMONTH] \THEYEAR}

\date{\specialdate\today}
\title{Checking Cheap Talk\thanks{We are grateful to Dirk Bergemann, V. Bhaskar, Tilman B\"orgers, Gabriel Carroll, Laura Doval, Matthew Elliott, Mira Frick, Sanjeev Goyal, Hans Peter Gr\"uner, Takakazu Honryo, Ryota Iijima, Navin Kartik, Dmitriy Knyazev, Volker Nocke, Martin Peitz, Hamid Sabourian, Larry Samuelson, Johannes Schneider, Nicolas Schutz, Andrzej Skrzypacz, Konrad Stahl, Takuo Sugaya, Emanuele Tarantino, Thomas Tr\"oger, and Juuso V\"alim\"aki as well as the audiences at Yale, Mannheim, Cambridge, and Birmingham for insightful comments and discussions.}}
\author{Ian Ball\thanks{MIT Department of Economics. Email: ianball@mit.edu} \hspace{6mm} Xin Gao\thanks{Department of Economics, University of Birmingham, UK. Email: x.gao.1@bham.ac.uk}}

\providecommand{\keywords}[1]{\textbf{Keywords:} #1}

\begin{document}

\maketitle

\begin{abstract}
We consider a sender--receiver game in which the receiver's action is binary and the sender's preferences are state-independent. The state is multidimensional. The receiver can select one dimension of the state to check (i.e., observe) before choosing his action. We identify a class of influential equilibria in which the sender's message reveals which components of the state are highest, and the receiver selects one of these components to check. In these equilibria, the sender's message guides the receiver to check positively selected attributes. We provide conditions under which these equilibria are sender-optimal, Pareto efficient, or receiver-optimal among all equilibria. 

\end{abstract}

\keywords{cheap talk, partial verification}

\textbf{JEL Classification Codes:} D82, D83

\newpage

\section{Introduction} 

Less informed agents often turn to biased experts for guidance. Since \cite{crawford1982strategic}, a large theoretical literature has demonstrated how cheap talk can influence a receiver's action, even when there is a conflict of interest between the sender and the receiver.  \cite{chakraborty2010persuasion} and \cite{lipnowski2020cheap} show that cheap talk can be influential even when the sender has \emph{state-independent} preferences. However, there remains an important case in which cheap talk cannot be influential: if the receiver's action is \emph{binary} and the sender strictly prefers the same action in every state. In this case,  the receiver must choose the same (mixed) action after every equilibrium message. Otherwise, the  sender would deviate by always sending the message that maximizes the probability of her preferred action. 

There are many applications in which the sender prefers the same action in every state. For example, a salesperson wants a shopper to buy her product, no matter the product's true quality. A job candidate wants to be hired, regardless of her skills. A politician wants a voter's support, regardless of whether her platform would benefit the voter. Communication in these settings is commonly observed, but it cannot be explained by the standard theory of cheap talk communication. 

In this paper, we show that costless communication can be influential in such settings if the receiver can gather additional information about the state after he receives the sender's message. For instance, the shopper can inspect some attributes of the product; the employer can test the candidate on certain tasks; and the voter can research some aspects of the politician's platform. Even though the receiver understands that the sender's message is intended to maximize the probability of the \emph{sender's} preferred action, the sender's message can influence \textit{which} information the receiver acquires. The salesperson can guide the shopper to inspect the best attributes of the product; the job candidate can steer the recruiter to test her strongest skills; and the politician can encourage the voter to research the most compelling parts of her platform. 

We consider a sender--receiver game, phrased in terms of our leading application of a seller and a buyer.  The state (of some product) is a vector of $N$ binary attributes. The state is drawn from a common prior that is exchangeable across the attributes. The receiver (he) chooses whether to buy the product at a fixed, exogenous price. The receiver's consumption utility is a strictly increasing function of the number of good attributes. The sender (she) strictly prefers the receiver to buy, regardless of the state. The timing is as follows. The sender observes the state and then sends a message to the receiver.  After seeing the message, the receiver chooses which attribute to ``check.'' The receiver observes whether the checked attribute is good or bad, and then decides whether to buy the product. We interpret the restriction on verification as a time or cognitive constraint, as in \citet{glazer2004optimal}. For example, a shopper may not have time to learn about every technical specification of a smartphone, and a recruiter does not have time to test every skill of a job candidate during a short interview. 


We construct a family of influential equilibria, parameterized by $k = 1, \ldots, N-1$. In the \emph{top-$k$ equilibrium}, the sender's message indicates which $k$ attributes are best (with ties broken by uniform randomization). The message does not reveal the ordering among these $k$ attributes. The receiver, upon seeing a message, learns that the $k$ attributes indicated by the message are more likely to be good than the other $N - k$ attributes, but the receiver's belief about the total number of good attributes remains unchanged. The receiver randomly selects one of the $k$ attributes to check. Then he buys if and only if that attribute is good. In these equilibria, the sender's message induces the receiver to check a positively selected attribute. In this way, communication can strictly benefit the sender by shaping the information that the receiver acquires. For example, the seller can direct the buyer to examine the best attributes of a smartphone, or a job candidate can suggest that her best skills be tested during the interview. 

\cref{res:top_single} characterizes the range of prices for which each top-$k$ equilibrium exists. This family of equilibria has an intuitive structure. As $k$ increases, the checked attribute becomes less positively selected. Thus, the checked attribute is less likely to be good, but if it is good, the receiver's belief shifts more. Therefore, as $k$ increases, the buying probability decreases, but the equilibrium can be sustained at higher prices.

Having characterized these top equilibria, we turn to results about all equilibria. It is difficult to directly characterize all equilibria. Our approach is to bound the space of outcomes that are incentive compatible for the sender. First, we consider buying probability vectors, which specify the average buying probability as a function of the number of good attributes. Because the receiver can check only one attribute, the sender's incentive constraints restrict how quickly the buying probability can increase in the number of good attributes (\cref{res:buying_prob}). We show that the buying probabilities induced by each top equilibrium satisfy these bounds with equality whenever the probabilities are interior. From each buying probability vector, we can compute the expected buying probability and the receiver's expected consumption value from buying. We define the \emph{IC-bounding polytope} to be the space of all such pairs that are induced by buying probability vectors satisfying the bounds from \cref{res:buying_prob}. This polytope provides an outer bound on the space of all incentive-compatible pairs. \cref{res:geometry} identifies geometric properties of this IC-bounding polytope. In particular,  its upper envelope is piecewise linear, and the vertices are induced by the top strategy profiles. Therefore, by analyzing the top equilibria, we can draw conclusions about \emph{all} equilibria. 

\cref{res:benefit} shows that the sender can strictly benefit from communication if and only if the sender strictly prefers one of the top equilibria to the no-communication outcome. So, analyzing the top equilibria is an easy way to determine whether the sender strictly prefers any equilibrium to the no-communication outcome.  

\cref{res:optimality} shows that the sender-preferred top equilibrium is Pareto efficient within the set of all equilibria. At the highest price at which the top-$k$ equilibrium exists, the top-$k$ equilibrium is also sender-optimal among all equilibria.  At this price, the top-$k$ equilibrium gives the receiver his lowest possible payoff among all equilibria---equal to what he can secure without communication by checking an arbitrary attribute. \cref{res:R_optimal} identifies the range of prices at which each top equilibrium is receiver-optimal among all equilibria. In fact, these equilibria achieve the receiver's maximum commitment payoff, so there is no value to commitment for the receiver. 

The main model assumes that the price is exogenous. Next, we allow the sender to choose the price (before observing the state). \cref{res:PE_pair} shows that the sender's payoff is maximized, over all price--equilibrium pairs, by some pair in which the equilibrium is either a top equilibrium or the no-communication outcome. These strategy profiles trace out the vertices of the upper envelope of the IC-bounding polytope. We construct an upper bound on the sender's payoff that is convex between adjacent vertices. Thus, this upper bound is maximized at some vertex, where it is achieved by an equilibrium. 

Finally,  we consider an extension in which the receiver can check $n$ of the $N$ attributes, where $1 < n < N$. We construct a natural analogue of each top-$k$ equilibrium. The sender's message reveals which $k$ attributes are best, and the receiver checks $\min\{ n, k\}$ attributes in that set. Then the receiver buys if and only if enough of the checked attributes are good. In \cref{res:multiple}, we characterize the interval of prices at which the top-$k$ equilibrium exists. As $k$ increases, this interval shifts upward, as in the baseline model with a single check. Even though the receiver's verification capacity is larger, the sender can still guide the receiver to acquire information about positively selected attributes.

The rest of the paper is organized as follows. After discussing related literature, we present  the model in \cref{mod}.  In \cref{checkone}, we characterize whether each top equilibrium exists. In \cref{sec:eq_bounds}, we bound the equilibrium buying probabilities and analyze the IC-bounding polytope. In \cref{sec:equilibrium_analysis}, we establish optimality properties of the top equilibria within the class of all equilibria. In \cref{sec:check_many}, we consider the extension in which the receiver can check multiple attributes. Section \ref{discussion} is the conclusion. Proofs omitted from the main text are in \cref{sec:proofs}.

\subsection{Related literature} \label{sec:lit}

Our model of verification is most closely related to the literature on screening with capacity-constrained verification. The key difference is that we focus on sender-optimal equilibria, rather than receiver-optimal mechanisms. Therefore, our analysis requires different techniques because we cannot restrict attention to direct, truthful communication. Specifically, \cite{glazer2004optimal} study a model in which the receiver, with commitment power, can check one component of the two-dimensional state before choosing a binary action.\footnote{In \cite{hancart2024optimal}, the receiver can select from a richer set of Blackwell experiments about the state. That paper characterizes whether it is optimal for the receiver to include a Blackwell-dominated experiment in the menu.} The sender has a strict, state-independent preference for one action over the other. The authors solve for the receiver-optimal mechanism and then show that this mechanism can be supported as an equilibrium, without receiver commitment.\footnote{\cite{GlazerRubinstein2006} solve for the receiver-optimal mechanism in a complementary problem in which the sender chooses which hard evidence to present to the receiver. They show that neither commitment nor randomization has value to the receiver. \cite{sher2011credibility} shows that this result extends beyond the binary setting, provided that the receiver's payoff is a concave transformation of the sender's payoff. More generally, \cite{sher2014persuasion} presents a linear program that characterizes  the ``dynamic'' persuasion problem from  \cite{glazer2004optimal} and then, with an additional integer constraint, characterizes the ``static'' persuasion problem from \cite{GlazerRubinstein2006}.} \cite{carroll2019strategic} generalize the setting of \cite{glazer2004optimal} by considering a multi-dimensional state and allowing the sender to have a more general utility function over the receiver's induced beliefs. They show that if the sender's indirect utility function is submodular in the state, then there exists a mechanism in which the receiver perfectly learns the state. This condition fails in our setting outside of edge cases.

Since \cite{crawford1982strategic}, the cheap talk literature has demonstrated the possibility of influential communication between players with partially aligned preferences. If the state and action spaces are multidimensional, and the players have \emph{state-dependent} preferences, then the sender can credibly reveal the ranking of the components of the state \citep{chakraborty2007comparative}. More recent work considers the extreme case in which the sender's preferences are \emph{state-independent}. \cite{chakraborty2010persuasion} and \cite{lipnowski2020cheap} show how communication can strictly benefit the sender when the sender's indirect payoff is not quasiconcave in the receiver's belief. With binary actions, however, cheap talk cannot be influential: in any equilibrium, all on-path messages must induce the same (mixed) action, and the receiver would be willing to take this action without any communication. In our seller--buyer game, without verification, cheap talk could not be influential. Verification introduces a new role for cheap talk---it can change the information that the receiver acquires.

The verification literature also studies multi-agent collective decision problems in which the principal can verify each agent's private information. The verification technology is modeled in different ways. Under \emph{costly verification}, the principal pays a cost for each agent's information that she verifies; see \cite{ben2014mechanisms}, who allocate a single item, and \cite{erlanson2020costly}, who study a binary collective decision. Under \emph{capacity-constrained verification}, verification is free, but the principal can verify at most a fixed number of agents; see \cite{erlanson2026optimal}, who allocate multiple items. Our model uses capacity-constrained verification in order to isolate how cheap talk can influence the receiver's choice of \emph{what} to verify. With costly verification, the receiver also chooses \emph{how much} to verify, trading off the value of additional information against its cost.

In many other models of verification, the receiver observes an \emph{exogenous} signal about the state. \cite{Kattwinkel2019} and \cite{silva2024information} give conditions under which the receiver, with commitment power, can benefit from communication, i.e., screening the sender. \cite{WekslerZik2025} give conditions under which communication can influence the receiver's action, when the sender has private information about the informativeness of the receiver's signal. In these papers, the sender's message influences the sensitivity of the receiver's action to his signal realization. In our model, by contrast, the sender's message influences \emph{which} signal the receiver observes. Finally,  communication equilibria have been analyzed in models where the receiver can (a) detect lies with an exogenous probability  \citep{balbuzanov2019lies,dziuda2018communication},  (b) choose to detect lies at a cost \citep{SadakaneTam2023}, or (c) choose to learn the state at a cost \citep{bijkerk2018words, venkatesh2024verifiable}.

In our model, the receiver can partially verify the state, but the sender's message is costless and unrestricted. An alternative modeling approach is taken in disclosure games
 \citep{milgrom1981good,grossman1981informational}. In that model, the sender can ``prove'' to the receiver any true statement about the state. Thus, messages have intrinsic meaning, and the true state determines which messages are feasible. Subsequent work imposes structure on the class of true statements that can be proven. In
 \cite{fishman1990optimal}, the sender can disclose one dimension of a multidimensional state. In \cite{dziuda2011strategic}, the sender can present verifiable ``arguments,'' and the number of arguments is unknown to the receiver.\footnote{A comprehensive survey on ``evidence'' in game theory and mechanism design can be found in \cite{ben2025evidence}.}

\section{Model} \label{mod}

There are two players: a sender (she) and a receiver (he). The sender privately observes the state $\th \in  \Th = \{ 0,1\}^N$. Assume $N \geq 2$. The state is drawn from a full-support prior $\pi \in \D ( \Th)$ that is exchangeable (i.e., invariant to permutations of the $N$ dimensions). The receiver chooses a binary action $a \in \{0,1\}$. In our leading interpretation, the sender is the seller of a product and the receiver is the buyer. The action  $a = 1$ denotes buying. Each component $\th_i$ represents a binary product attribute.\footnote{For example, the components could be interpreted as the outcomes of binary product tests (pass or fail), which are independently and identically distributed conditional on the (unobserved) product quality.} The utilities of the sender and receiver are given by 
\[
	u_S( a, \th) = p a, \qquad u_R( a, \th) =  (  v(\th) - p)a ,
\]
where $p > 0$ is a fixed, exogenous price, and $v(\th)$ is the receiver's consumption utility from the product in state $\th$. We assume that the utility function $v \colon \Th \to \R_+$ is strictly increasing and symmetric across the attributes. That is, $v$ can be expressed as a strictly increasing function of the number of good attributes, which we denote by $\abs{\th}  \coloneqq \th_1 + \cdots + \th_N$. For $j = 0, \ldots, N$, let $\pi_j$ denote the probability that $\abs{\th} = j$ and let $v_j$ denote the consumption utility $v(\th)$ in any state $\th$ with $\abs{\th} = j$. 

The receiver has a limited capacity to learn about the product.  Following \citet{glazer2004optimal}, we assume that the receiver can check one component of the state $\th$.\footnote{In  \cref{sec:check_many}, we consider the case in which the receiver can check multiple components.}  Checking attribute $i$ perfectly reveals $\th_i$. Since this check is costless, we assume that the receiver always checks exactly one attribute. We interpret this capacity constraint as a hard limit on the receiver's time or cognitive resources. It can also capture an institutional constraint such as a platform-review process in which a set number of items is checked. The timing is as follows. The sender observes the state and sends a costless message to the receiver. The receiver sees the message and then chooses which of the $N$ attributes to check. The receiver observes whether the checked attribute is good ($1$) or  bad  ($0$), and then chooses whether to buy the product.

Next, we define strategies. Let $M$ be a (sufficiently rich) message space.\footnote{Formally, suppose that $|M| \geq 2^N$, so that messages can index every subset of attributes. If $M$ is infinite, then we assume that $M$ is a Polish space and that all maps are Borel-measurable.}  A \emph{messaging strategy} for the sender is a function $m \colon \Theta \to \D (M)$, which assigns to each state a distribution over messages in $M$. Let $[N] =\{ 1, \ldots, N \}$. A strategy for the receiver is a pair $(c,b)$ consisting of (i) a \textit{checking strategy} $c \colon M \to \Delta([N])$, which assigns to each message a probability distribution over attributes, and (ii) a \textit{buying strategy}
\[
	b \colon M \times [N] \times \{0,1\} \to [0,1],
\]
which specifies the probability that the receiver buys ($a = 1$) as a function of the message, the attribute that is checked, and the value of that checked attribute. The solution concept is Nash equilibrium, which we simply call  \emph{equilibrium}. Each top equilibrium that we construct below can be supported as a sequential equilibrium; see \cref{rem:seq_eq}. Formally, working with Nash equilibrium strengthens our results about the optimality of the top equilibria among all equilibria. 

\section{Constructing the top equilibria} \label{checkone}

In this section, we construct a family of influential equilibria, dubbed the top equilibria. 


\subsection{Benchmark: No verification}

As a benchmark, suppose that the receiver cannot check any component of the state, as in a standard cheap talk game. Since the sender always wants the receiver to buy, cheap talk cannot induce the receiver to take any action that he would not take in the absence of communication. Indeed, the sender's incentive constraints imply that all equilibrium messages must induce the same buying probability. If $p  < \E [ v(\th)]$, then in every equilibrium, the receiver buys the product. If $p > \E [ v(\th)]$, then in every equilibrium, the receiver does not buy the product. If $p = \E[ v(\th)]$, then the receiver is indifferent between buying and not buying. For each $q \in [0,1]$, there is an equilibrium in which the receiver buys  with probability $q$.

\subsection{Top equilibria} \label{sec:no_comm}

We now return to the main model with verification. We first consider what happens if the sender does not communicate with the receiver. Since the prior is exchangeable, the receiver is indifferent between checking each of the $N$ attributes. Whichever attribute he checks, he observes whether that attribute is good or bad, so by exchangeability, his updated expectation of his consumption utility is either
\[
\bar{\nu} \coloneqq \E [ v(\th) \mid \th_1= 1]
\quad
\text{or}
\quad
\ubar{\nu} \coloneqq \E [ v(\th) \mid \th_1 = 0].
\]
Observe that $\ubar{\nu} < \bar{\nu}$.  We define the no-communication outcome, assuming that the receiver breaks ties in the sender's favor. If $p \leq \ubar{\nu}$, the receiver always buys. If $\ubar{\nu} < p \leq \bar{\nu}$, the receiver buys if and only if the checked attribute is good. If $p > \bar{\nu}$, the receiver never buys. Of course, this no-communication outcome can be supported as a babbling equilibrium. 

We now construct a family of influential equilibria, indexed by $k  = 1, \ldots, N-1$. In the top-$k$ equilibrium, the sender's message reveals which $k$ attributes are best, with ties broken uniformly. The receiver randomly selects one of these $k$ ``recommended'' attributes to check. Then the receiver buys if and only if the checked attribute is good. For example, suppose that $N=3$. Consider the top-$2$ equilibrium. If $\theta=(1,0,0)$, then with probability $1/2$ the sender's message indicates that ``attributes 1 and 2 are the two best attributes,'' and with probability $1/2$ the sender's message indicates that ``attributes 1 and 3 are the two best attributes.'' To be sure, messages have no intrinsic meaning; their interpretation is determined endogenously by the sender's equilibrium strategy.

Formally, for $k = 1, \ldots, N-1$, the \emph{top-$k$ strategy profile} $(m_k; c_k, b_k)$ is defined as follows. Let $\mathcal{P}_k$ denote the collection of all $k$-element subsets of $[N]$. Since $M$ is sufficiently rich, we may identify a subset of $M$ with $\cup_{k=1}^{N-1} \mathcal{P}_k$. For each $A \in \mathcal{P}_k$, let $\abs{\th_A} =  \textstyle \sum_{i \in A} \th_i$. That is, $\abs{\th_A}$ counts the number of good attributes in $A$. For each state $\th$, let $m_k (\th)$ be the uniform distribution over $\argmax\nolimits_{A \in \mathcal{P}_k }\,  \abs{ \th_A}$. For each $A \in \mathcal{P}_k$, let $c_k (A)$ be the  uniform distribution over $A$. For some fixed $A_0 \in \PP_k$, let $c_k(A') = c_k(A_0)$ for all $A' \in M \setminus \PP_k$. Finally, let $b_k ( A, i, z) = z$ for each $(A,i,z) \in M \times [N] \times \{0,1\}$. That is, the receiver buys if and only if the attribute that he checks is good; recall that, by exchangeability, the receiver's belief about the total number of good attributes is unaffected by the sender's message. If the top-$k$ strategy profile is an equilibrium, then we call it the \emph{top-$k$ equilibrium}, and we say that the top-$k$ equilibrium exists. 

Our first result provides a necessary and sufficient condition for the existence of the top-$k$ equilibrium. To state this condition, we introduce notation for the signals that the receiver can obtain, given the sender's top-$k$ strategy, by checking either a recommended or an unrecommended attribute. For $k = 1, \ldots, N-1$, let $T^k$ ($B^{N-k}$) denote the value of an independent uniform draw from the top $k$ (bottom $N-k$) attributes, where these top (bottom) attributes are determined by uniform tie-breaking. In state $\th$, no matter how ties are broken, we know that among the top $k$ attributes, $\abs{\th} \wedge k$ are good, and among the bottom $N-k$ attributes, $( \abs{\th} - k)_+$ are good.\footnote{Throughout this paper, $\wedge$ denotes the minimum operator, and we let $x_+ = \max\{ x, 0\}$.} Therefore, given the state $\th$,  the random variables $T^k$ and $B^{N-k}$ are  conditionally independent Bernoulli random variables, with
\[
	\P (T^k = 1 \mid \th) = \frac{ \abs{\th} \wedge k}{k}, \qquad \P (B^{N-k}  =1 \mid \th) = \frac{( \abs{\th} -k)_+}{N-k}.
\]
Note that the conditional distributions depend only on the number $|\th|$ of good attributes. For $k = 1, \ldots, N-1$, define the thresholds 
\[
	\ubar{p}_k = \E [ v(\th) \mid T^k = 0], 
	\qquad
	\bar{p}_k = \E [v(\theta) \mid (T^k, B^{N-k})=(1,0)].
\]
We will interpret these thresholds after stating the equilibrium condition. 

\begin{thm}[Top-$k$ equilibrium condition]
\label{res:top_single}
For $k =1, \ldots, N-1$, the top-$k$ equilibrium exists if and only if
\[
 \ubar{p}_k \leq  p \leq  \bar{p}_k.
\]
Moreover, these price thresholds are ordered as follows:
\[
	v_0 = \ubar{p}_1 < \cdots < \ubar{p}_{N-1}  < \ubar{\nu} <  \bar{p}_1 < \cdots < \bar{p}_{N-1} < \bar{\nu}.
\]
\end{thm}

For each $k$, the top-$k$ equilibrium exists if and only if the price $p$ lies in the nondegenerate interval $[\ubar{p}_k, \bar{p}_k]$. The lower bound $p \geq \ubar{p}_k$ ensures that if the receiver checks one of the $k$ recommended attributes and sees that it is bad, then he finds it optimal not to buy.  The upper bound $p \leq \bar{p}_k$ ensures that the receiver finds it optimal to check one of the recommended attributes, rather than one of the unrecommended attributes. To get intuition for this, suppose that the receiver deviates by checking one of the unrecommended attributes and then buying if and only if that attribute is good. Every recommended attribute is weakly better than every unrecommended attribute, so under this deviation, the receiver's buying decision differs from the equilibrium in the event that the checked unrecommended attribute is bad, but the recommended attribute that the receiver would have checked is good. This corresponds to the event that $(T^k, B^{N-k}) = (1,0)$. The inequality $p \leq \bar{p}_k$ ensures that the receiver weakly prefers buying to not buying, conditional on this event. 


\cref{fig:top_equilibria} illustrates the buying probability under the top equilibria and the no-communication outcome in a numerical example. In this example,  $N = 4$ and $v(\th) = \abs{\th}$. The state distribution is such that $\abs{\th}$ is equally likely to take the values $0, \ldots, 4$.\footnote{Thus, $\P (T^k = 1) = (1/5) \textstyle\sum_{j=0}^{4} (j \wedge k)/k$. This expression equals $8/10$, $7/10$, $6/10$ for $k = 1,2,3$ respectively.} For each $k =1, 2, 3$, we plot the buying probability under the top-$k$ equilibrium over the range of prices at which the top-$k$ equilibrium exists. As $k$ increases, the buying probability shifts down, and the price range $[\ubar{p}_k, \bar{p}_k]$ shifts to the right. Intuitively, as $k$ increases, the checked attribute becomes less positively selected relative to a randomly chosen attribute. Thus, the checked attribute is less likely to be good, but  if it is good, the receiver's belief shifts more, allowing the equilibrium to be sustained at higher prices. The dashed line indicates the buying probability without communication: for $p \leq \ubar{\nu}$, the receiver always buys; for $\ubar{\nu} < p \leq \bar{\nu}$, the receiver buys if and only if the checked attribute is good.\footnote{For $p > \bar{\nu}$, the receiver never buys. This part of the graph is omitted from the plot.} Over the shaded price range $(\ubar{\nu}, \bar{p}_3]$, there exists a top equilibrium that the sender strictly prefers to the no-communication outcome. 

\begin{figure}
	\centering
	\begin{tikzpicture}
		\begin{axis}[
			xmin = 0, xmax = 3.5,
			ymin = 0.4, ymax = 1.1,
			axis lines = middle,
			xtick = {1, 7/3, 3},
			xticklabels = {$\ubar{\nu}$, $\bar{p}_3$, $\bar{\nu}$},
			ytick = {0.5,1},
			yticklabels = {0.5,1},
			clip = false, 
			xlabel = {$p$},
			xlabel style={at=(current axis.right of origin), anchor=west},
			ylabel = {$\P ( a = 1)$},
			ylabel style={at=(current axis.north west), anchor=south}, 
            axis on top
			]

        \fill[gray!25]
	       (axis cs:1,0.4) rectangle (axis cs:7/3,0.43);
            
			\addplot[thick, Blue] coordinates { (0,4/5) (5/3,4/5)} node[right]{$k = 1$};
			\addplot[thick, Orange] coordinates { (1/3, 7/10) (2,7/10)} node[right]{$k = 2$};
			\addplot[thick, Green] coordinates { (2/3,3/5) (7/3,3/5)} node[right]{$k = 3$};
			\addplot[thick, dashed] coordinates { (0,1) (1,1)};
			\addplot[thick, dashed] coordinates { (1,1/2) (3,1/2)} node[right]{no comm};
		\end{axis}
	\end{tikzpicture}
	\caption{Buying probability under top equilibria and no-communication outcome}
	\label{fig:top_equilibria}
\end{figure}

\begin{rem}[Sequential equilibrium] \label{rem:seq_eq} For $k = 1, \ldots, N-1$, whenever the top-$k$ strategy profile is a (Nash) equilibrium, the same outcome can be induced by a sequential equilibrium.\footnote{The standard definition of sequential equilibrium applies to finite games, so for this result (and its analog in \cref{rem:seq_eq_n}), we assume that $M$ is finite.} Here is a sketch of the argument. For each $A \in \mathcal{P}_k$, the receiver's beliefs after observing $A$ are pinned down by Bayes' rule. For each $A' \in M \setminus  \mathcal{P}_k$, the receiver can form the same belief as if he had observed $A_0$; these beliefs are consistent. Each of the receiver's beliefs preserves uncertainty about the value of every attribute, so whatever attribute the receiver checks, and whatever value he observes, his beliefs can be updated by Bayes' rule. Whenever the receiver deviates to check an unrecommended attribute, his buying decision can be modified so that it is sequentially rational. This modification changes behavior only after a deviation by the receiver, so it does not affect the sender's incentives. 
\end{rem}

\begin{rem}[Exchangeability and symmetry] \label{rem:exch_sym} Our construction of the top-$k$ equilibrium uses the exchangeability of the state distribution $\pi$ and the symmetry of the receiver's valuation function $v$. The importance of these assumptions depends on $k$. In the top-$1$ equilibrium, the receiver does not mix, so as long as $\ubar{p}_1 < p < \bar{p}_1$, the receiver's obedience conditions hold strictly. Therefore, the top-$1$ strategy profile would remain an equilibrium under all sufficiently small perturbations of $\pi$ and $v$. For example, starting from $v(\th) = | \th|$, we could allow $v(\th) = \a_1 \th_1 + \cdots + \a_N \th_N$, provided that $\a_1, \ldots, \a_N$ are all sufficiently close to $1$. For $k > 1$, the top-$k$ equilibrium requires the receiver to mix, which relies on the exact indifference supplied by exchangeability and symmetry.  
\end{rem}

Hereafter, we extend the definition of the top-$k$ strategy profile to $k=0$ and $k = N$. 
In each case, the sender conveys no information. In the top-$0$ strategy profile, the receiver always buys. In the top-$N$ strategy profile, the receiver checks an arbitrary attribute and buys if and only if the checked attribute is good.\footnote{In each case, there are multiple strategy profiles with this property, but all such strategy profiles are payoff equivalent.} Let $\bar{p}_0 = \ubar{\nu}$ and $\bar{p}_N = \bar{\nu}$. For $k =0$ and $k = N$, note that $\bar{p}_k$ is the highest price at which the top-$k$ strategy profile is an equilibrium. 

To be sure, there can be other equilibria that are not mixtures between top strategy profiles. For example, at some prices there is an equilibrium in which the sender reveals the top $k$ attributes among a fixed proper subset of attributes; the receiver randomly selects one of these $k$ ``recommended'' attributes to check and then buys if and only if the checked attribute is good.\footnote{Alternatively, if $N$ is even, then for $k < N/2$, there can be equilibria in which the sender indicates the top $k$ among the first $N/2$ attributes, and also the top $k$ among the next $N/2$ attributes, and the receiver randomly checks one of these $2k$ ``recommended'' attributes.}  It is difficult to characterize all equilibria. We instead bound the space of all equilibrium outcomes. 

\section{Bounding incentive-compatible outcomes} \label{sec:eq_bounds}

In order to prove results about all equilibria, we establish outer bounds on the space of outcomes that satisfy the sender's incentive constraints. Then we show that the top equilibria induce extreme points of this bounding set. The main welfare results in \cref{sec:equilibrium_analysis} will follow directly from the technical analysis in this section. 

\subsection{Bounding incentive-compatible buying probabilities} \label{sec:q_def}

Under each strategy profile, the expected payoffs for the sender and the receiver are pinned down by the average buying probability conditional on each realization of $\abs{\th}$. We record these probabilities in the buying probability vector $q = (q_0, \ldots, q_N)$, where $q_j$ denotes the average buying probability conditional on the event that $\abs{\th} = j$. We use the sender's incentive constraints to derive bounds on these buying probabilities.

\begin{lem}[Bounding buying probabilities] \label{res:buying_prob}
For any strategy profile in which the sender is playing a best response, the induced buying probabilities $q_0, \ldots, q_N$ satisfy
\begin{equation} \label{eq:buying_ineq}
	q_{j-1} \geq \frac{j-1}{j} q_{j},  \qquad j = 1, \ldots, N.
\end{equation}
\end{lem}

The sender's incentive constraints limit how steeply the buying probability can increase as a function of the number $\abs{\th}$ of good attributes. The maximal steepness reflects the precision of the verification technology. Indeed, without verification, the buying probability function must be flat. With perfect verification of the full state, the buying probability function is not restricted by the sender's incentive constraints. To be clear, the condition in \eqref{eq:buying_ineq} is necessary but not sufficient for incentive compatibility. In particular, the sender's incentive constraints also limit how steeply the buying probability can \emph{decrease} as a function of the number of good attributes. 

To prove \cref{res:buying_prob}, we fix an arbitrary strategy profile in which the sender is playing a best response. We analyze a special class of sender deviations. Consider some type $\th'$ with $\abs{\th'} =  j \geq 1$. Choose a dimension $i$ such that $\th_i' = 1$, and consider the state $\th = \th'- e_i$, where  $e_i$ is the $i$-th standard basis vector. Type $\th$ can mimic the (possibly random) message sent by type $\th'$. If the receiver checks any dimension other than dimension $i$, then he cannot distinguish type $\th$ from type $\th'$, so the receiver makes the same buying decision as if the sender's true type were $\th'$. Since type $\th$ cannot profit from mimicking type $\th'$, we obtain a lower bound on the state-$\th$ buying probability that depends on the state-$\th'$ buying probability. We then average over such states $\th$ and $\th'$ to obtain the bound in \eqref{eq:buying_ineq}. 

Let $Q$ be the set of all vectors $q$ in $[0,1]^{N+1}$ that satisfy \eqref{eq:buying_ineq}. The set $Q$ is a polytope. We introduce notation for the buying probabilities
under the top strategy profiles. For $k = 1, \ldots, N-1$, let 
\begin{equation}\label{eq:q_top_k}
	q_j^k =  \P (T^k = 1 \mid \abs{\th} = j) = \frac{ j \wedge k}{k}, \qquad  j= 0, \ldots, N.
\end{equation}
The vector $q^k$ satisfies \eqref{eq:buying_ineq} with equality for each $j$ such that $q_{j-1}^{k} < 1$. Each vector $q^k$ is an extreme point of $Q$ since the lower bound on $q_{j-1}$ in \eqref{eq:buying_ineq} holds with equality for $j = 1, \ldots, k$ and $q_j^k = 1$ for $j = k, \ldots, N$. 

We introduce notation for a few other buying probability vectors. 
Define the vector $q^N$ by extending the definition in \eqref{eq:q_top_k} to $k = N$.  Set $q^0 = \bm{1}$ and $q^{N+1} = \bm{0}$. The vectors $q^0, q^{N}, q^{N+1}$ are also extreme points of $Q$. These vectors satisfy the componentwise inequalities $q^{N+1} \leq q^{N} \leq \cdots \leq q^{0}$.

\subsection{Geometry of the IC-bounding polytope}

The sender's and receiver's payoffs from a buying probability vector $q$ are determined by two statistics: the ex-ante buying probability $x(q)$ and the receiver's expected utility from consumption $y(q)$, defined by
\begin{equation} \label{eq:x_def}
	x(q)  = \sum_{j=0}^{N} \pi_j q_j \quad \text{and} \quad y(q)  = \sum_{j=0}^{N} \pi_j q_j v_j.
\end{equation}
The sender's payoff is $p x(q)$ and the receiver's payoff is $y(q) - p x(q)$. 

It is more convenient to work with these two statistics than the full $(N+1)$-vector of  buying probabilities. Define the \emph{IC-bounding polytope} by 
\[
\XX  = \{ (x(q), y(q)): q \in Q \}.
\]
The set $\XX$ is indeed a polytope since it is the image of the polytope $Q$ under the linear transformation $q \mapsto (x(q), y(q))$.  We work with the set $\XX$ rather than with the corresponding payoff pairs because the set $\XX$ does not depend on the price $p$, and we will vary $p$ in some of our analysis below. Since $Q$ contains every constant vector in $[0,1]^{N+1}$, the projection of $\XX$ onto its first component is $[0,1]$. Define the function $\hat{y} \colon [0,1] \to \R$ by 
\[
    \hat{y}(x) =  \max \{ y : (x,y) \in \XX\}.
\]
The graph of $\hat{y}$ is the upper envelope of $\XX$. 

Next, we describe points in $\XX$ induced by the top strategy profiles. Recall the vectors $q^{0}, \ldots, q^{N+1}$ defined in \cref{sec:q_def}. For $k = 0, \ldots, N+1$, let
\[
    (x_k, y_k) = ( x( q^k), y(q^k)).
\]
Thus, 
\[
    0 = x_{N+1} < x_N < \cdots < x_1 < x_0 = 1.
\]
For each $k = 0, \ldots, N$, let 
\[
    \b_k =  \frac{ y_k -  y_{k+1}}{x_k - x_{k+1}}.
\]
Thus, $\b_k$ is the slope of the line segment connecting $(x_{k+1}, y_{k+1})$ to $(x_k, y_k)$. Economically, $\b_k$ is the receiver's average consumption value for the marginal purchases that take place under $q^k$ but not under $q^{k+1}$. 

\begin{figure}
\centering
\begin{tikzpicture}
    \begin{axis}[
        width=300pt,
        xmin=0, xmax=1.1,
        ymin=0, ymax=2.1,
        axis lines=middle,
        xtick={1},
        xticklabels={1},
        ytick={1,2},
        yticklabels={1,2},
        clip=false,
        xlabel={$x(q)$},
        xlabel style={
            at=(current axis.right of origin),
            anchor=west
        },
        ylabel={$y(q)$},
        ylabel style={
            at=(current axis.north west),
            anchor=south
        },
    ]

        \addplot[
            mark=none,
            draw=none,
            fill=gray!50,
            opacity=0.5
        ] coordinates {
            (0,0)
            (0.2,0)
            (0.4,0.2)
            (0.6,0.6)
            (0.8,1.2)
            (1,2)
            (0.8,2)
            (0.7,1.9)
            (0.6,26/15)
            (0.5,1.5)
        } -- cycle;

        \addplot[
            mark=*,
            mark size=1.5pt
        ] coordinates {
            (1,2)
            (4/5,2)
            (7/10,19/10)
            (3/5,26/15)
            (1/2,3/2)
            (0,0)
        };

        \path (axis cs:4/5,2) --
            node[midway, below] {$\b_0$}
            (axis cs:1,2);

        \path (axis cs:7/10,19/10) --
            node[midway, below, xshift = 4pt, yshift = 1pt] {$\b_1$}
            (axis cs:4/5,2);

        \path (axis cs:3/5,26/15) --
            node[midway, below, xshift = 4pt, yshift = 1pt] {$\b_2$}
            (axis cs:7/10,19/10);

        \path (axis cs:1/2,3/2) --
            node[midway, below, xshift = 4pt, yshift = 2pt] {$\b_3$}
            (axis cs:3/5,26/15);

        \path (axis cs:0,0) --
            node[midway, right=1pt] {$\b_4$}
            (axis cs:1/2,3/2);

        \node at (axis cs:1,2)
            [anchor=south] {$0$};

        \node at (axis cs:4/5,2)
            [anchor=south] {$1$};

        \node at (axis cs:7/10,19/10)
            [anchor=south] {$2$};

        \node at (axis cs:3/5,26/15)
            [anchor=south] {$3$};

        \node at (axis cs:1/2,3/2)
            [anchor=south] {$4$};
        
        \node at (axis cs:0,0)
            [anchor=south, yshift =2pt, xshift = -5pt] {$5$};

        \node at (axis cs:0.55,0.9)
            [anchor=south] {$\XX$};

    \end{axis}
\end{tikzpicture}
\caption{IC-bounding polytope and its upper envelope}
\label{fig:upper_envelope}
\end{figure}

\cref{fig:upper_envelope} plots the set $\XX$ in the same numerical example with $N =4$ used in \cref{fig:top_equilibria}. For $k = 0, \ldots, N+1$, each point $(x_k, y_k)$ is labeled as $k$. The upper envelope of $\XX$ is the union of the line segments connecting the adjacent vectors in the sequence $(x_{5}, y_{5}), \ldots, (x_{0}, y_{0})$. This geometric property holds in general, as we show next.

\begin{lem}[Geometry of IC-bounding polytope]  \label{res:geometry} The following hold.
\begin{enumerate}[label = (\roman*), ref = \roman*]
    \item \label{it:UE} The upper envelope of $\XX$ equals  $\cup_{k=0}^{N} [ (x_{k+1}, y_{k+1}),(x_k, y_k)]$.
    \item \label{it:prices} For $k = 0, \ldots, N-1$, we have 
    \[
        \bar{p}_k = \frac{y_k - y_N}{x_k - x_N}.
    \]
    \item \label{it:price_slope_comparison} For $k = 1, \ldots, N-1$, we have 
    \[
        \ubar{p}_k \leq \b_{k-1} < \b_k \leq \bar{p}_k,
    \]
    where the last inequality holds with equality if and only if $k = N-1$. Also, 
    \[
    \b_0 = v_0 < \ubar{\nu} < \b_{N-1} < \b_N = \bar{\nu}.
    \]
    
\end{enumerate}
\end{lem}

In part~\ref{it:UE}, the images of the buying probability vectors $q^0, \ldots, q^{N+1}$ lie on the upper envelope of $\XX$ because in these vectors, the buying probability increases as steeply as is permitted by \eqref{eq:buying_ineq} whenever it is below the boundary value of $1$. Part~\ref{it:prices} holds because for each $k = 0, \ldots, N-1$, at price $\bar{p}_k$ the receiver is indifferent between the top-$k$ strategy and deviating by checking a random attribute and buying if and only if that checked attribute is good. Geometrically, $\bar{p}_k$ is the slope of the line segment between $(x_k, y_k)$ and $(x_N, y_N)$. Finally, consider part~\ref{it:price_slope_comparison}. The inequality between the slopes holds because, as $k$ decreases, the marginal purchases occur when fewer attributes are good. The inequality with the upper price threshold follows from comparing the slopes of the secant lines to the slopes of the line segments in the upper envelope. The inequality $\ubar{p}_k \leq \b_{k-1}$ is proven from a direct likelihood-ratio comparison. 


The following welfare results (Theorems~\ref{res:benefit}--\ref{res:PE_pair}) will be established by analyzing the geometry of the IC-bounding polytope and its upper envelope. 

\section{Equilibrium welfare} \label{sec:equilibrium_analysis}

In this section, we use the top equilibrium construction and the geometric properties of the IC-bounding polytope to obtain welfare results about all equilibria. First, we characterize whether the sender can strictly benefit from communication. Then we give conditions under which each top equilibrium is Pareto efficient, sender-optimal, or receiver-optimal among all equilibria. We also show that the receiver gets no value from commitment.

\subsection{Benefiting from communication}
 
We say that the sender \emph{benefits from communication} if there exists an equilibrium that the sender strictly prefers to the no-communication outcome (defined in \cref{sec:no_comm}).

\begin{thm}[Benefiting from communication]  \label{res:benefit} The sender benefits from communication if and only if
	\[
	\ubar{\nu} < p \leq \bar{p}_{N-1}.
	\]
\end{thm}
Thus, some equilibrium benefits the sender if and only if some top equilibrium does. The price range $(\ubar{\nu}, \bar{p}_{N - 1}]$ is shaded in \cref{fig:top_equilibria}, where $N-1 = 3$.

\begin{proof}[Proof of Theorem~\ref{res:benefit}]
By \cref{res:top_single}, we have $\ubar{p}_{N-1} < \ubar{\nu} < \bar{p}_{N-1} < \bar{\nu}$. We argue by cases.
\begin{enumerate}
    \item $p \leq \ubar{\nu}$. Without communication, the receiver always buys, so the sender cannot do any better. 
    \item $\ubar{\nu} < p \leq \bar{p}_{N-1}$. Without communication, the receiver checks an arbitrary attribute and buys if and only if it is good. The top-$(N-1)$ equilibrium exists and induces a strictly higher buying probability.
	\item $\bar{p}_{N-1} < p \leq \bar{\nu}$. The no-communication outcome is $(x_N, y_N)$. For each  point $(x,y)$ in $\XX$, if $x > x_N$, then
    \begin{equation} \label{eq:dev_N}
        y - p x \leq \hat{y}(x) - p x < y_N - p x_N,
    \end{equation}
    where the last inequality follows from \cref{res:geometry}: the line segment from $(x_N, y_N)$ to $(x,\hat{y}(x))$ has slope at most $\b_{N-1}$, and $\b_{N-1} = \bar{p}_{N-1} <  p$. By \eqref{eq:dev_N}, the point $(x,y)$ cannot be induced by an equilibrium, since the receiver can ignore the sender and guarantee $y_N - p x_N$.
    \item $p > \bar{\nu}$. The no-communication outcome is $(0,0)$. For each point $(x,y)$ in $\XX$, if $x > 0$, then 
    \begin{equation} \label{eq:dev_0}
    y - p x \leq \hat{y}(x) - px < 0,
    \end{equation}
    where the last inequality follows from \cref{res:geometry}: the line segment from $(0,0)$ to $(x,\hat{y}(x))$ has slope at most $\b_{N}$, and $\b_{N} = \bar{\nu} <  p$. By \eqref{eq:dev_0}, the point $(x,y)$ cannot be induced by an equilibrium, since the receiver can ignore the sender and never buy to guarantee payoff $0$. \qedhere
\end{enumerate}
\end{proof}

\cref{res:benefit} shows that the sender's incentive constraints can prevent mutually beneficial communication. For prices $p$ slightly above $\bar{\nu}$, if there is no communication, then the receiver never buys. The receiver would be willing to buy upon seeing that one of the \emph{worst} attributes was good. But in an influential equilibrium, the sender cannot induce the receiver to check a negatively selected attribute. If the receiver believes that he is checking one of the worst attributes, then the sender would strictly prefer to guide him to one of the best attributes instead.

\subsection{Optimality of top equilibria}

We say that an equilibrium is \emph{Pareto efficient} if it is not Pareto dominated by any other equilibrium. An equilibrium is \emph{sender-optimal} if no other equilibrium gives the sender a strictly higher payoff. 

\begin{thm}[Pareto efficiency and sender optimality] \label{res:optimality} For $k  = 1, \ldots, N-1$, the top-$k$ equilibrium is Pareto efficient if $\b_{k-1} < p \leq \bar{p}_{k}$ and sender-optimal if $p = \bar{p}_k$.
\end{thm}

By \cref{res:geometry}.\ref{it:price_slope_comparison}, we have $\ubar{p}_k \leq \b_{k-1}$, so over the price range $(\b_{k-1}, \bar{p}_k]$, the top-$k$ equilibrium exists. \cref{res:geometry}.\ref{it:price_slope_comparison} implies that $\b_{k-1} < \bar{p}_{k-1}$ for $k = 2, \ldots, N-1$, and $\b_{0} = \ubar{p}_1$ (by \cref{res:top_single}). Therefore, if $p > v_0$, \cref{res:optimality} implies that whenever there exists a top equilibrium, the sender-preferred top equilibrium is Pareto efficient. 

\begin{proof}[Proof of Theorem~\ref{res:optimality}]

Fix  $k \in \{ 1, \ldots, N -1\}$. Fix a price $p \in (\b_{k-1}, \bar{p}_k]$. For every point $(x,y)$ in $\XX$, if $x > x_k$, then
\begin{equation} \label{eq:p_k_bound}
y - p x \leq \hat{y} (x) - px < y_k - p x_k,
\end{equation}
where the last inequality holds as follows. By \cref{res:geometry}.\ref{it:UE}, the slope of the line segment from $(x_k, y_k)$ to $(x, \hat{y}(x))$ is a weighted average of $\b_{k-1}, \ldots, \b_0$. By \cref{res:geometry}.\ref{it:price_slope_comparison}, this weighted average is at most $\b_{k-1}$, which is strictly less than $p$. 

First we prove Pareto efficiency. Fix $(x,y) \in \XX$. If $x = x_k$, then $y \leq \hat{y}(x_k) = y_k$. If $x > x_k$, then $y - p x < y_k - p x_k$ by \eqref{eq:p_k_bound}. Next, we check sender-optimality. Let $p = \bar{p}_k$. By \cref{res:geometry}.\ref{it:prices}, we have  $y_k  - p x_k = y_N - p x_N$. Substituting this equality into \eqref{eq:p_k_bound}, and using the fact that the receiver can guarantee a payoff of $y_N - p x_N$ by ignoring the sender, 
we conclude that no equilibrium can induce a point $(x,y)$ with $x > x_k$. 
\end{proof}

So far, we have focused on the sender's payoffs from the top equilibria. Now we turn to the receiver's payoffs. Given the price $p > 0$, define the receiver's commitment payoff to be the receiver's maximum payoff over all strategy profiles $(m;c,b)$ in which $m$ is a best response to $(c,b)$. 

\begin{thm}[Receiver optimality] \label{res:R_optimal} The receiver's optimal commitment payoff is achieved at an equilibrium. For $k = 1, \ldots, N$, it is achieved by the top-$k$ strategy profile if $\b_{k-1} \leq p \leq \b_k$. It is achieved by always buying if $p \leq \b_0 = v_0$ and by never buying if $p \geq \b_N = \bar{\nu}$. 
\end{thm}

\begin{proof}[Proof of Theorem~\ref{res:R_optimal}]
Every strategy profile $(m;c,b)$ in which $m$ is a best response to $(c,b)$ induces an outcome $(x,y)$ in $\XX$. Maximize the receiver's payoff function, $y - p x$, over $\XX$. A maximizer exists since the function is continuous and $\XX$ is compact. Since this function is strictly increasing in $y$, every maximizer lies on the upper envelope of $\XX$. By \cref{res:geometry}.\ref{it:UE}, the upper envelope of $\XX$ consists of line segments with slopes $\b_{N}  > \cdots > \b_0$. Therefore, 
\[
    \argmax_{(x,y) \in \XX} (y - p x) = 
    \begin{cases}
        \{(x_0, y_0)\} &\text{if}~ p < \b_0, \\
        [(x_{k+1},y_{k+1}),(x_k,y_k)] &\text{if}~ p = \b_k~ \text{for some}~ k \in \{0, \ldots, N\}, \\
        \{(x_k,y_k)\} &\text{if}~ \b_{k-1} < p < \b_{k}~ \text{for some}~k \in \{ 1, \ldots, N\}, \\
        \{(x_{N+1}, y_{N+1})\} &\text{if}~ p > \b_N.
    \end{cases}
\]
It remains to show that each maximizer in $\{ (x_k, y_k)\}_{k=0}^{N+1}$ is induced by some equilibrium. In view of \cref{res:top_single}, this follows from \cref{res:geometry}.\ref{it:price_slope_comparison}. For $k = 1, \ldots, N-1$, we have $\ubar{p}_k \leq \b_{k-1} < \b_k \leq \bar{p}_k$. The edge cases $k = 0, N, N+1$ follow from the inequality
\[
    \b_0 = v_0 < \ubar{\nu} < \b_{N-1} < \b_N = \bar{\nu}. \qedhere
\]
\end{proof}


This proof can be easily visualized on \cref{fig:upper_envelope} since the level curves of the linear map $(x,y) \mapsto y - p x$ have slope $p$. 

Finally, we compare  \cref{res:R_optimal} with \cref{res:optimality}. For each $k = 1, \ldots, N-1$, the top-$k$ equilibrium is Pareto efficient over the price range $(\b_{k-1}, \bar{p}_k]$. By \cref{res:geometry}.\ref{it:price_slope_comparison}, we have $\b_{k-1} < \b_k \leq \bar{p}_k$. At the high price $p = \bar{p}_k$, the top-$k$ equilibrium is sender-optimal.  Over the low price range $(\b_{k-1}, \b_k]$, the top-$k$ equilibrium is receiver-optimal. For $k < N-1$, the top-$k$ equilibrium is not receiver-optimal over $(\b_{k}, \bar{p}_k]$ because there the receiver prefers the top-$(k+1)$ strategy profile.  

\subsection{Endogenous pricing}
Up to this point, we have assumed that the price $p$ is exogenous. Here, we assume that the sender can choose the price  (before observing the state). We maximize the sender's payoff over all price--equilibrium pairs. 

\begin{thm}[Optimal price and equilibrium] \label{res:PE_pair}  There exists $k^\ast \in \{ 0, \ldots, N\}$ such that the sender's payoff is maximized, over all price--equilibrium pairs, by the price $p = \bar{p}_{k^\ast}$ and the top-$k^\ast$ strategy profile.
\end{thm}

One implication of \cref{res:PE_pair} is that in order to find an optimal price and equilibrium, it suffices to optimize over finitely many points. To prove \cref{res:PE_pair}, we maximize the sender's payoff $px$ over vectors $(p,x,y)$, where $(x,y)$ is in $\XX$ and $(p,x,y)$ satisfies necessary conditions for receiver obedience---the receiver's payoff is at least as high as what he can guarantee by ignoring the sender and checking an arbitrary attribute. We show that the solution is achieved by a vertex on the upper envelope of $\XX$.

The optimal price--equilibrium pair depends on the state distribution and the utility function $v$. Consider the simple example with $N = 2$ and $v(\th) = \abs{\th}$.  It can be verified that the top-$1$ equilibrium is optimal if $\pi_1 \geq 2 \pi_2$ and the top-$2$ equilibrium is optimal if $\pi_1 \leq 2 \pi_2$. For the calculations, see \cref{sec:optimum_calculation}. When $\abs{\th} = 1$, the receiver buys with certainty in the top-$1$ equilibrium but only with probability $1/2$ in the top-$2$ equilibrium. Thus, as $\pi_1/\pi_2$ increases, the buying probability ratio between the top-$1$ equilibrium and the top-$2$ equilibrium increases. The price ratio $\bar{p}_1/ \bar{p}_2$ also increases because the receiver, upon seeing that a randomly chosen attribute is $1$, assigns lower probability to the event $\abs{\th} = 2$.

\section{Checking multiple attributes} \label{sec:check_many}

In the baseline model, the receiver can check exactly one attribute. In this section, we assume that the receiver can simultaneously check $n$ of the $N$ attributes, where $1 < n < N$. Since these $n$ checks are costless, we assume that the receiver always checks exactly $n$ attributes. We rule out the case of perfect verification $(n = N)$: if the receiver can learn the state perfectly, then there is no role for communication. 

First, we define strategies in this extended model. As in the baseline model, a messaging strategy for the sender is a function $m \colon \Th \to \D (M)$. Let $\mathcal{P}_n$ denote the collection of all $n$-element subsets of $[N]$. A strategy for the receiver is a pair $(c,b)$, consisting of (i) a \emph{checking strategy} $c \colon M \to \D ( \PP_n)$, specifying which $n$ attributes are checked, and (ii) a \emph{buying strategy} 
\[
	b \colon M \times \PP_n \times \{ 0,1\}^n \to [0,1],
\]
which specifies the probability that the receiver buys ($a = 1$) as a function of the message, the set of attributes that are checked, and the values of the checked attributes.\footnote{In $\{0,1\}^n$, the $i$-th component is the value of the checked attribute with the $i$-th smallest index.}

We now extend the notion of a top-$k$ equilibrium to this setting. The top-$k$ messaging strategy $m_k$ for the sender is defined as in the baseline model. The receiver's top-$k$ strategy is as follows. The receiver uniformly selects $n \wedge k$ of the recommended attributes and $(n -k)_+$ of the unrecommended attributes to check. Then the receiver buys if and only if the number of good attributes among the recommended attributes that he checks weakly exceeds some threshold, which is between $1$ and $n \wedge k$. The values of the unrecommended attributes do not affect the receiver's buying decision.\footnote{If the receiver were allowed to check \emph{at most} $n$ attributes, then the receiver could equivalently check only $n \wedge k$ of the recommended attributes.} As in the baseline model, the receiver treats any off-path message as if he received the same fixed on-path message. We say that the top-$k$ equilibrium exists if such a strategy profile is an equilibrium.\footnote{In order to speak of \emph{the} top-$k$ equilibrium, we assume that if there are multiple thresholds in $\{ 1, \ldots, n \wedge k\}$ that make the receiver's strategy a best response, the smallest such threshold is chosen.}

As in the baseline model, we give a necessary and sufficient condition for the existence of the top-$k$ equilibrium. To state the condition, we generalize our notation. For $k = 1, \ldots, N$, let $T_n^k$ denote the number of good attributes in a uniform sample of $n \wedge k$ of the top $k$ attributes. Let $B_n^{N-k}$ denote the number of good attributes in a uniform sample of $n \wedge (N-k)$ of the bottom $N-k$ attributes. In each case, the top and bottom attributes are determined by uniform tie-breaking. For $k = 1, \ldots, N-1$, define the thresholds
\[
		\ubar{p}_{k,n} =\E [v(\theta) \mid  T_n^k = 0], 
		\qquad
			\bar{p}_{k,n} =\E[v(\theta) \mid (T_{n}^k, B^{N-k}_{n})=(n \wedge k,0)].
\]
With $n =1$, these definitions coincide with the definitions from the main model: $\ubar{p}_{k,1} = \ubar{p}_k$ and $\bar{p}_{k,1} = \bar{p}_k$. 

\begin{thm}[Top-$k$ equilibrium condition with multiple attributes checked]
\label{res:multiple}
Suppose $1 < n < N$. For $k = 1, \ldots, N-1$, the top-$k$ equilibrium exists if and only if
\[
	\ubar{p}_{k,n} \leq p \leq \bar{p}_{k,n}.
\]
Moreover, $\ubar{p}_{k,n} < \bar{p}_{k,n}$ for each $k$, and the price thresholds are ordered as follows:
\[
	v_0 = \ubar{p}_{1,n} = \cdots = \ubar{p}_{n,n} < \cdots < \ubar{p}_{N-1,n} \quad \text{and} \quad
	\bar{p}_{1,n} < \cdots < \bar{p}_{N-1,n}.
\]
\end{thm}

\begin{rem}[Sequential equilibrium] \label{rem:seq_eq_n}  For $k = 1, \ldots, N-1$, whenever the top-$k$ strategy profile is a (Nash) equilibrium, the same outcome can be induced by a sequential equilibrium. The argument is similar to that in \cref{rem:seq_eq} for the baseline model, except that for some $n$-element subsets of attributes, some attribute values are inconsistent with the sender's top-$k$ strategy. In this case, the receiver can form the most pessimistic beliefs consistent with the observation, and then the receiver's buying decision can be modified so that it is sequentially rational. This modification will not introduce profitable deviations for the sender.\footnote{Indeed, by deviating, the sender induces the receiver to check weakly fewer good attributes (in the sense of first-order stochastic dominance) and to form weakly more pessimistic beliefs conditional on the number of good attributes observed.}
\end{rem}

For $k = 1, \ldots, N-1$, the top-$k$ equilibrium exists if and only if the price $p$ lies in the nondegenerate interval $[\ubar{p}_{k,n}, \bar{p}_{k,n}]$. This equilibrium price interval shifts up as $k$ increases, just as in the baseline model with $n=1$. Intuitively, as the sender recommends more attributes, the sample of recommended attributes becomes less positively selected, so the equilibrium can be sustained at higher prices. 

When the receiver can check multiple attributes, the set of deviations is much richer. To prove \cref{res:multiple}, the trick is to compute the receiver's best response to the sender's top-$k$ strategy in a hypothetical specification in which the receiver can acquire more information---checking $n \wedge k$ of the $k$ recommended attributes and $n \wedge (N-k)$  of the $N-k$ unrecommended attributes. Thus, the receiver effectively observes $(T_n^k, B_n^{N-k})$. Here, we sketch the proof in the special case that $n\leq k$ and $n\leq N-k$.  The random vector $(T_n^k, B_n^{N-k})$ can only take the values 
\[
	(0,0),(1,0),\ldots,(n,0),(n,1),\ldots,(n,n).
\]
These vectors are totally ordered by the pointwise order. A standard statistical argument shows that the conditional expectation of $v(\th)$ is strictly increasing along this sequence. Therefore, in the hypothetical specification, the receiver has a best response to the sender's top-$k$ strategy of the following form: buy if and only if $(T_n^k, B_n^{N-k})$ exceeds some threshold vector. In order to implement this best response, the receiver does \emph{not} need to observe both $T_n^k$ and $B_n^{N-k}$. For $0 \leq t \leq n$, we have: $(T_n^k, B_n^{N-k}) \geq (t,0)$ if and only if $T_n^k \geq t$. For $0 < b \leq n$, we have: $(T_n^k, B_n^{N-k}) \geq (n,b)$ if and only if $B_n^{N-k} \geq b$. Thus, every threshold best response in the hypothetical specification is feasible, hence optimal, in the original specification. The upper bound $p \leq \bar{p}_{k,n}$ ensures that there exists an optimal threshold of the form $(t,0)$ for some $0 \leq t \leq n$. The lower bound $p \geq \ubar{p}_{k,n}$  ensures that there is such a threshold with $t > 0$, so that the receiver does not always buy. 

As in the baseline model, the price thresholds satisfy an intuitive order. As $k$ increases, the recommended attributes become less positively selected, so the equilibrium can be sustained at higher prices. On the other hand, the effect on the buying probability is generally ambiguous because the buying threshold also changes.  For $1 < n <  N$ and $1 \leq k \leq N-1$, and  $\ubar{p}_{k,n} \leq p \leq \bar{p}_{k,n}$, define the buying threshold by
\[
    t^\ast (p,k,n) = \min \Set{ t \in \{ 1, \ldots, n \wedge k \}  : \E [ v(\th) \mid (T_n^k, B_n^{N-k}) = (t,0) ] \geq p  }.
\]
In the top-$k$ equilibrium, the receiver buys if and only if at least $t^\ast (p,k,n)$ of the checked recommended attributes are good. In the baseline model with $n=1$, this threshold is always $1$. 

\begin{thm}[Buying threshold] \label{res:buying_threshold}
For $1 < n \leq k \leq N-1$ and $\ubar{p}_{k,n} \leq p \leq \bar{p}_{k,n}$, the buying threshold $t^\ast (p,k,n)$ is weakly decreasing in $k$ and weakly increasing in $p$ and $n$.
\end{thm}

Naturally, the threshold is weakly increasing in the price $p$. As $k$ increases, the checked attributes become less positively selected, so the threshold weakly decreases. As the number of checked attributes $n$ increases, the threshold weakly increases.

When multiple attributes can be verified, the sender's incentive constraints impose weaker restrictions on the buying probabilities. Since the threshold $t^\ast (p,k,n)$ varies with $p$, the buying probability vectors induced by the top strategy profiles are not necessarily extreme points within the set of vectors satisfying these restrictions. Therefore, it is difficult to derive general conclusions about all equilibria from analyzing only the top equilibria.

\begin{rem}[Adaptive checking] \label{rem:adaptive} We have assumed that the receiver selects all $n$ attributes before seeing any of their values. If the receiver could instead check the attributes adaptively, based on the realizations of previously checked attributes, then our proof of \cref{res:multiple} would still go through. Conditional on the sender's top-$k$ message, the $k$ recommended attributes are exchangeable and the $N- k$ unrecommended attributes are exchangeable. Therefore, checking $n$ attributes adaptively is still less informative than checking $n \wedge k$ of the recommended attributes and $n \wedge (N- k)$ of the unrecommended attributes.
\end{rem}

\section{Conclusion}
\label{discussion}

In this paper, we study how a sender with state-independent preferences can benefit from costless communication when the receiver can partially verify the state. We identify a natural class of equilibria, the \emph{top equilibria}, in which the sender guides the receiver to learn about the best components of the state. Because these equilibria induce extreme points of the set of incentive-compatible outcomes, this construction allows us to draw conclusions about all equilibria. We give conditions under which each top equilibrium is Pareto efficient, sender-optimal, or receiver-optimal. 

Following \citet{glazer2004optimal}, we have assumed that checking attributes is costless up to some checking capacity, and prohibitively costly beyond that capacity. This captures a hard constraint on the receiver's time or cognitive resources.  An interesting direction for future work is incorporating a more flexible model of information acquisition, in which the receiver chooses how much to inspect as well as what to inspect.

\appendix

\section{Omitted proofs} \label{sec:proofs}

\subsection{Proof of Theorem \ref{res:top_single}}

Fix $k \in [N-1]$.  Under the top-$k$ strategy profile, it is clear that the sender is playing a best response to the receiver's strategy since recommending a top-$k$ set maximizes the buying probability. We characterize whether the receiver is playing a best response to the sender's strategy. Suppose that the sender uses the top-$k$ messaging strategy. Further,  suppose counterfactually that the receiver could acquire more information (in the sense of Blackwell) by checking one recommended attribute and one unrecommended attribute.  By exchangeability, a sufficient statistic for the receiver's observation is $(T^k, B^{N-k})$. The support of this random variable is $\{ (0,0), (1,0), (1,1)\}$. By \cref{res:favorability} in \cref{sec:favorability}, 
\begin{equation}
\begin{aligned}
    	\E [ v ( \th) \mid (T^k, B^{N-k}) = (0,0) ] 
        &< 	\E [ v ( \th) \mid (T^k, B^{N-k}) = (1,0) ] \\
        &< 	\E [ v ( \th) \mid (T^k, B^{N-k}) = (1,1) ].
\end{aligned}
\end{equation}
 Thus, the receiver has a best response that takes one of the following threshold forms: (i) never buy, (ii) buy if and only if $T^k =1$, (iii) buy if and only if $B^{N- k} = 1$, and (iv) always buy. Each of these strategies is implementable in the original game. Therefore, (ii) is a best response in the original game if and only if (ii) is a best response in the counterfactual game. And this holds if and only if $\ubar{p}_k \leq p \leq \bar{p}_k$.

It remains to establish the inequality between the price thresholds. By definition, $\ubar{p}_1 = v_0$. By \cref{res:favorability_thresholds}.\ref{it:t}, we have $\ubar{p}_1 < \cdots < \ubar{p}_{N-1} < \ubar{\nu}$. By \cref{res:favorability_thresholds}.\ref{it:TB}, we have $\ubar{\nu} < \bar{p}_1 < \cdots < \bar{p}_{N-1} < \bar{\nu}$.

\subsection{Proof of Lemma~\ref{res:buying_prob}}

Fix a strategy profile $(m; c, b)$ in which $m$ is a best response to $(c,b)$. For each $\th \in \{0,1\}^N$, let $q(\th)$ be the buying probability in state $\th$. For each $\th$ and each $i$ for which $\th_i = 1$, let
\[
d_i (\th)  = \E_{m' \sim m(\th)} \Brac{ c_i (m') b(m', i, 1)}.
\]
Thus, $d_i(\th)$ is the probability that the receiver checks attribute $i$ and buys. We have
\begin{equation} \label{eq:q_bound}
    \sum_{i : \th_i = 1} d_i (\th) \leq q(\th). 
\end{equation}

Now we prove \eqref{eq:buying_ineq}. This inequality is trivial if $j = 1$, so we fix $j \in \{2, \ldots, N\}$. Choose $\th$ with $|\th|  = j$.  For each $i$ with $\th_i = 1$, we have
\[
    q( \th - e_i) \geq q(\th) - d_i (\th).
\]
Averaging over the indices $i$ with $\th_i=1$ gives
\begin{equation*}
      \frac{1}{j} \sum_{i: \th_i= 1} q( \th - e_i) 
      \geq  q(\th) -  \frac{1}{j} \sum_{i: \th_i= 1} d_i(\th) \geq \frac{j - 1}{j} q(\th),
\end{equation*}
where the last inequality uses \eqref{eq:q_bound}. Average over all $\tbinom{N}{j}$ states $\th$ with $\abs{\th} = j$. For each state $\th'$ with $|\th'| = j-1$, there exist $N-j + 1$ states $\th$ with $|\th| = j$ that differ from $\th'$ in exactly one component, so the weight on each state $\th'$ is
\[
    \frac{N - j +1}{j \binom{N}{j}}= \frac{1}{\binom{N}{j-1}}.
\]
Since $\pi$ is exchangeable, we conclude that
\[
q_{j-1} \geq \frac{j-1}{j} q_j.
\]

\subsection{Proof of Lemma~\ref{res:geometry}}

\paragraph{Part~\ref{it:UE}} First observe that the function $\hat{y}$ is well-defined on $[0,1]$ since $Q$ contains every constant vector in $[0,1]^{N+1}$. Since $\cup_{k=0}^{N} [ (x_{k+1}, y_{k+1}),(x_k, y_k)]$ is the graph of a function on $[0,1]$, it suffices to prove that the graph of $\hat{y}$ is a subset of $\cup_{k=0}^{N} [ (x_{k+1}, y_{k+1}),(x_k, y_k)]$. Fix $x \in [0,1]$. Choose $q \in Q$ such that $(x(q), y(q)) = (x, \hat{y} (x))$. We claim that for all $k \in \{ 0, \ldots, N-1\}$, we have $q_k = \tfrac{k}{k+1} q_{k+1}$ or $q_{k+1} = 1$. Otherwise, there exists $k \in \{ 0, \ldots, N-1\}$ such that $q_k > \tfrac{k}{k+1} q_{k+1}$ and $q_{k+1} < 1$. We perturb the vector $q$ by scaling down $q_0, \ldots, q_{k}$ and increasing $q_{k +1}$ in such a way that the expected buying probability is preserved. Given $\e > 0$, define $\tilde{q} =  (\tilde{q}_0, \ldots, \tilde{q}_N)$ by 
\[
	\tilde{q}_j 
	= \begin{cases} 
	(1- \e) q_j &\text{if}~j \leq k,\\
		q_j +  \e \sum_{\ell = 0}^{k} (\pi_\ell/\pi_{k+1}) q_\ell &\text{if}~j = k +1,\\
		q_j &\text{if}~j >  k +1.
	\end{cases}
\]
If $\e$ is chosen sufficiently small, then $\tilde{q}_k > \tfrac{k}{k+1} \tilde{q}_{k+1}$ and $\tilde{q}_{k+1} < 1$. In this case, the vector $\tilde{q}$ is in $Q$. By construction, $x(\tilde{q}) = x(q)$ and $y ( \tilde{q}) > y (q)$, contrary to the definition of $\hat{y}(x)$. 

Now using the claim, we show that $q$ is in $\cup_{k=0}^{N} [q^k, q^{k+1}]$. If $q = \bm{1}$, then $q$ is in $[q^0, q^{1}]$, so we may assume that $q \neq \bm{1}$. Let $k$ be the largest index $j$ such that $q_j < 1$. If $k = N$, then $q_j = \tfrac{j}{j+1} q_{j+1}$ for all $j = 0, \ldots, N-1$, so $q = q_N q^N \in [q^N, q^{N+1}]$. If $k  < N$, then $q_j = 1$ for all $j > k$. By the definition of $Q$, we have $q_k \geq \tfrac{k}{k+1} q_{k+1}$. If $k = 0$, then $q$ is in $[q^0, q^{k+1}]$. If $k > 0$, then applying the claim repeatedly, we conclude that $q_j = (j/k)q_k$ for all $j = k-1, k-2, \ldots, 0$.  Therefore, $q$ is in $[q^{k}, q^{k+ 1}]$.

\paragraph{Part~\ref{it:prices}}  Let $[\cdot ]$ denote the indicator function for the predicate that it encloses. Fix $k \in \{1, \ldots, N-1\}$. For $j = 0, \ldots, N$, observe that
	\begin{equation*}
		\begin{aligned}
			q_j^k - q_j^N  &= 	\frac{j  (N-k)} { k N}  [ j < k] + \frac{N - j}{N}  [j \geq k] \\
			&=  \frac{N-k}{N}  \cdot \P \Paren{ (T^k, B^{N-k}) = (1,0)  \mid \abs{\th} = j }.
		\end{aligned}
	\end{equation*}
	Therefore, 
	\begin{equation*}
		\begin{aligned}
    \frac{y_k - y_N}{x_k - x_N}
    &= \frac{\sum_{j=0}^{N} \pi_j v_j (q_j^k - q_j^N)}{\sum_{j=0}^{N} \pi_j (q_j^k - q_j^N)} \\
    &= \E \Brac{ v(\th) \mid (T^k, B^{N-k}) = (1,0) } \\
    &= \bar{p}_k.
    \end{aligned}
    \end{equation*}

For $k = 0$, we have
\begin{equation*}
		\begin{aligned}
    \frac{y_0 - y_N}{x_0 - x_N}
    &= \frac{\sum_{j=0}^{N} \pi_j v_j (1 - j/N)}{\sum_{j=0}^{N} \pi_j (1 - j/N)} \\
    &= \E \Brac{ v(\th) \mid \th_1 = 0} \\
    &= \ubar{\nu} \\
    &= \bar{p}_0.
    \end{aligned}
    \end{equation*}

\paragraph{Part~\ref{it:price_slope_comparison}} First we compare the slopes. We have $\b_0 = v_0$, and for $k = 1, \ldots, N$,
\begin{equation} \label{eq:beta_def}
	\b_k = \frac{ \sum_{j=0}^{k} \pi_j j v_j}{\sum_{j=0}^{k} \pi_j j}.
\end{equation}
We have $v_0 < \cdots < v_N$. Thus, $\b_0 = v_0 < v_1 = \b_1$. For $k = 2, \ldots, N$, since $\b_{k-1} \leq v_{k-1} < v_{k}$ and $\b_{k}$ is a strict convex combination of $\b_{k-1}$ and $v_{k}$, we have $\b_{k-1} < \b_{k}$. By \eqref{eq:beta_def}, we have $\b_N = \bar{\nu}$. 

Now we compare these slopes with the upper price thresholds. Fix $k \in \{1, \ldots, N-2\}$. Since $\bar{p}_k$ is a strict convex combination of $\b_k, \ldots, \b_{N-1}$, and we have just shown that $\b_k < \cdots < \b_{N-1}$, it follows that $\b_k < \bar{p}_k$. By definition, we have $\bar{p}_{N-1} = \b_{N-1}$. 

Finally, we compare these slopes with the lower price thresholds. For $k = 1$, we have $\ubar{p}_1 = v_0 = \b_0$. Fix $k \in \{2, \ldots, N-1\}$. We have 
\[
    \ubar{p}_k = \frac{ \sum_{j=0}^{k-1}  \pi_j (k- j)  v_j}{\sum_{j=0}^{k-1}  \pi_j(k - j)} 
    \quad
    \text{and}
    \quad
    \b_{k-1} = \frac{\sum_{j=0}^{k-1} \pi_j j v_j}{\sum_{j=0}^{k-1} \pi_j j}. 
\]
These are conditional expectations, and we get $\ubar{p}_k \leq \b_{k-1}$ by likelihood ratio dominance since the ratio $(k - j)/j$ is decreasing in $j$ over $\{0, \ldots, k - 1\}$. Taking $k = N$, the same argument gives $\ubar{\nu} < \b_{N-1}$.

\subsection{Proof of Theorem~\ref{res:PE_pair}}

It suffices to show that for any price $p \geq 0$ and any point $(x,y) \in \XX$ satisfying 
\begin{equation} \label{eq:eq_bounds}
    y - p x \geq 0 \quad \text{and} \quad y - p x \geq y_N - p x_N,
\end{equation}
we have $p x \leq \textstyle\max_{k = 0, \ldots, N} \bar{p}_k x_k$. To see why this is sufficient, note that any price--equilibrium pair induces a point in $\XX$ satisfying \eqref{eq:eq_bounds} since the receiver can ignore the sender's message and then either never buy or implement the top-$N$ profile. 

Now we complete the proof. Fix $p \geq 0$. For each $(x,y) \in \XX$ satisfying \eqref{eq:eq_bounds}, if $x \leq x_N$, then 
\[
    p x \leq y \leq \hat{y} (x) = \b_N x \leq \bar{p}_N x_N,
\]
where the last inequality follows from \cref{res:geometry}. For each $(x,y) \in \XX$ satisfying \eqref{eq:eq_bounds}, if $x > x_N$, then
\begin{equation} \label{eq:N_bound}
    px \leq \frac{ y - y_N}{x - x_N} x \leq \frac{ \hat{y}(x) - y_N}{x - x_N} x.
\end{equation}
Denote the right side by $R(x)$. It remains to bound the function $R$ over $(x_N, 1]$. Recall that $x_N < x_{N-1} < \cdots < x_1 < x_0 = 1$. For $x \in (x_N, x_{N-1}]$, we have
\[
    R(x) = \b_{N-1} x \leq \b_{N-1} x_{N-1} = \bar{p}_{N-1} x_{N-1},
\]
where the last equality follows from \cref{res:geometry}.\ref{it:price_slope_comparison}. It suffices to show that for $k = 0, \ldots, N-2$, the function $R$ is convex over $[x_{k+1}, x_k]$ for then the maximum of $R$ over $[x_{k+1}, x_k]$ is achieved at an endpoint, hence
\[
    R(x) \leq \max \{ R(x_k), R( x_{k+1}) \} = \max\{ \bar{p}_k x_k, \bar{p}_{k+1} x_{k+1} \},
\]
where the final equality follows from \cref{res:geometry}. Fix $k \in \{0, \ldots, N-2\}$. For every $x \in [x_{k+1}, x_k]$, we have
\[
    \hat{y}(x) - y_N = \b_k (x -x_N) + ( \bar{p}_k -\b_k) (  x_k- x_N),
\]
so
\[
    R(x) = \b_k x + ( \bar{p}_k -\b_k) (  x_k- x_N) \frac{ x}{x - x_N}.
\]
The map $x \mapsto x/(x - x_N)$ is convex and the coefficient $( \bar{p}_k -\b_k) (x_k- x_N)$ is nonnegative by \cref{res:geometry}.\ref{it:price_slope_comparison}. Thus, $R$ is convex over $[x_{k+1}, x_k]$.

\subsection{Example for Theorem~\ref{res:PE_pair}} \label{sec:optimum_calculation}


Consider the simple example with $N = 2$ and $v(\th) = \abs{\th}$. The upper price thresholds are given by
\begin{align*}
			\bar{p}_0 &= \E  \Brac{ \abs{\th} \mid \th_1  = 0 } = \frac{\pi_1/2}{\pi_0 + \pi_1/2},  \\
			\bar{p}_1 &= \E \Brac{\abs{\th} \mid (T^1, B^1) = (1,0)} = 1,\\
			\bar{p}_2 &= \E  \Brac{ \abs{\th} \mid \th_1  = 1 } = \frac{ \pi_1/2 + 2 \pi_2}{\pi_1/2 + \pi_2}.
\end{align*}
The associated maximum revenues are
\begin{align*}
			 \bar{p}_0 &= \frac{\pi_1/2}{\pi_0 + \pi_1/2}, \\
			 \bar{p}_1 ( \pi_1 + \pi_2)  &= \pi_1 + \pi_2, \\
		 \bar{p}_2 ( \pi_1/2 + \pi_2) &= \pi_1/2 + 2 \pi_2.
\end{align*}
	Since $\pi_0 = 1 - \pi_1 - \pi_2$, we have $\bar{p}_1 ( \pi_1 + \pi_2) \geq \bar{p}_0$ if and only if 
	\[
	\pi_1 + \pi_2  \geq \pi_1/2 + (\pi_1/2 + \pi_2) ( \pi_1  + \pi_2),
	\]
	or equivalently,
	\[
	\pi_1/ 2 + \pi_2 \geq (\pi_1/2 + \pi_2) ( \pi_1  + \pi_2),
	\]
	which holds because $\pi_1 + \pi_2 \leq 1$. Therefore, either the top-$1$ equilibrium or the top-$2$ equilibrium is optimal. The top-$1$ equilibrium is optimal if $\pi_1 \geq 2 \pi_2$ and the top-$2$ equilibrium is optimal if $\pi_1 \leq 2 \pi_2$. Note that 
	\[
		  \frac{ \bar{p}_1 (\pi_1 + \pi_2) }{ \bar{p}_2 ( \pi_1/2 + \pi_2)} = \frac{\pi_1/2 + \pi_2}{\pi_1/2 + 2 \pi_2} \cdot \frac{ \pi_1 + \pi_2}{\pi_1/2 + \pi_2}.
	\]
	So the price ratio and the buying probability ratio are both increasing in $\pi_1/\pi_2$.

\subsection{Favorability order} \label{sec:favorability}

We are interested in the distribution of $\abs{\th}$, conditional on events involving $(T_{n}^k, B_n^{N-k})$. For any such event $E$, let
\[
\supp E  = \{ j \in \{0, \ldots, N\}: \P ( E \mid \abs{\th} = j) > 0 \}.
\]
Note that this notion is different from the standard notion of the support of a random variable. Consider events $E$ and $E'$ with positive probability. We say that $E$ is \emph{weakly more favorable} than $E'$ if the ratio 
\begin{equation} \label{eq:ratio}
	\frac{	\P ( E  \mid  \abs{\th} = j )}{\P ( E'  \mid  \abs{\th} = j )}
\end{equation}
is weakly increasing in $j$ over $\supp E \cup \supp E'$; we use the convention that $a/0= \infty$ for $ a> 0$. If, further, the ratio in  \eqref{eq:ratio} is not constant over $\supp E \cup \supp E'$, then we say that $E$ is \emph{strictly more favorable} than $E'$.\footnote{Our terminology is inspired by \cite{milgrom1981good}. Unlike in his definition, we do not assume that the events have full support.  Our formal definition mirrors the definition of the likelihood ratio order in \citet[1.C.1, p.~42]{SS}.} If $E$ is strictly more favorable than $E'$, then  $\E [ v(\th) \mid E ] > \E [ v(\th) \mid E']$, by standard results about the likelihood ratio order.\footnote{ If $E$ is weakly more favorable than $E'$, then the distribution of $\abs{\th}$, conditional on $E$, likelihood ratio dominates the distribution of $\abs{\th}$, conditional on $E'$, in the sense of \citet[1.C.1, p.~42]{SS}. Moreover, likelihood ratio dominance implies first-order stochastic dominance \citep[Theorem 1.C.1, p.~43]{SS}.} Let 
\[
    S(n,k) = \{ (0,0), \ldots, (n \wedge k, 0), (n \wedge k, 1), \ldots, (n \wedge k, n\wedge (N-k)) \}.
\]


\begin{lem}[Favorability]  \label{res:favorability} 
For each $n,k \in [N-1]$, the favorability of the event $(T_{n}^k, B_n^{N-k}) = (t,b) $ is strictly increasing in $(t,b)$ over the range $S(n,k)$, with respect to the product order.
\end{lem}

For the next result, extend the definition of $B_n^{N-k}$ to $B_n^N$, and adopt the convention that $T_n^0 = B_n^0 = 0$ for $n = 1, \ldots, N$.

\begin{lem}[Favorability across equilibria] \label{res:favorability_thresholds} The following hold. 
	\begin{enumerate}[label = (\roman*), ref = \roman*]
		\item \label{it:t} For $1 \leq n \leq k \leq N$ and $0 \leq t \leq n$, the favorability of the event $T_n^k = t$ is strictly increasing in $k$ and strictly decreasing in $n$. 
		\item \label{it:TB} Fix $n \in \{1, \ldots, N-1\}$. The favorability of the event $(T_n^k, B_n^{N-k}) = (n \wedge k, 0)$ is strictly increasing in $k$ over $\{ 0, \ldots, N\}$. 

	\end{enumerate}
\end{lem}

These lemmas are proven in \cref{sec:proof_fav1,sec:proof_fav2}.

\subsection{Proof of Theorem~\ref{res:multiple}} \label{sec:proof_multiple}

Fix $k \in [N-1]$.  Under the top-$k$ strategy profile, it is clear that the sender is playing a best response to the receiver's strategy since recommending a top-$k$ set maximizes the buying probability. We characterize whether the receiver is playing a best response to the sender's strategy. Suppose that the sender uses the top-$k$ messaging strategy. Further,  suppose counterfactually that the receiver could acquire more information (in the sense of Blackwell) by checking $n \wedge k$ of the recommended attributes and also $n \wedge (N-k)$ of the unrecommended attributes. By exchangeability, a sufficient statistic for the receiver's observation is $(T_n^k, B_n^{N-k})$. The support of this random variable is $S(n,k)$. Note that this set is totally ordered by the product order. By \cref{res:favorability}, the conditional expectation
\[
	\E [ v ( \th) \mid (T_n^k, B_n^{N-k}) = (t,b) ]
\]
is strictly increasing in $(t,b)$ over $S(n,k)$. Thus, in the counterfactual problem, the receiver has a best response that takes one of the following threshold forms: (i) never buy; (ii) buy if and only if $T_n^k \geq t^\ast$ for some $t^\ast \in \{1, \ldots, n \wedge k \}$; (iii) buy if and only if  $B_n^{N-k} \geq b^\ast$ for some $b^\ast \in  \{ 1, \ldots, n \wedge (N-k)\}$; (iv) always buy. Therefore, (ii) is a best response in the original game if and only if (ii) is a best response in the counterfactual game. And this holds if and only if $\ubar{p}_{k,n} \leq p \leq \bar{p}_{k,n}$.

Now we turn to the price thresholds. For each $k$, the inequality $\ubar{p}_{k,n} < \bar{p}_{k,n}$ follows from \cref{res:favorability} since $T_n^k = 0$ if and only if $(T_n^k, B_n^{N-k}) = (0,0)$. The ordering of the lower price thresholds follows from \cref{res:favorability_thresholds}.\ref{it:t} whenever $k \geq n$. Observe that $\ubar{p}_{k,n} = v_0$ for $k \leq n$. The ordering of the upper price thresholds follows from \cref{res:favorability_thresholds}.\ref{it:TB}.

\subsection{Proof of  Lemma~\ref{res:favorability}} \label{sec:proof_fav1}

We first introduce notation for the hypergeometric distribution with a sample of size $s$ from a population of $S$ units of which $g$ are good. The associated density is given by
\[
    h_s^S ( x | g )  = \frac{ \binom{g}{x} \binom{S - g}{s - x}}{\binom{S}{s}},
\]
where we use the convention that $\textstyle\binom{b}{a} = 0$ unless $0 \leq a \leq b$.  For $0 \leq x < s$, consider the ratio
\[
    \frac{h_s^S(x+1 | g)}{h_s^S(x | g)}
    =
    \frac{g-x}{x+1} \cdot
    \frac{s-x}{S-g-s+x+1}.
\]
Whenever the denominator is strictly positive, this ratio is strictly increasing in $g$. We will use this observation below.

Now we turn to the main proof. Fix $n,k \in \{1, \ldots, N-1\}$. For each $(t,b) \in S(n,k)$ and $0 \leq j \leq N$,  let
\begin{equation*}
\begin{aligned}
f( t, b | j) 
&= \P  \Paren{ (T_{n}^k, B_n^{N-k}) = (t,b) \mid \abs{\th} = j} \\
&=  h_{n \wedge k}^{k} ( t | j \wedge k) h^{N-k}_{n \wedge (N-k)} ( b| (j - k)_+).
\end{aligned}
\end{equation*}
Observe that the support of the event $(T_n^k, B_n^{N- k}) = (t,b)$ is given by 
\[
\begin{cases}
    [t, k - (n \wedge k) + t] &\text{if}~t < n \wedge k, \\
    [n \wedge k, N - ( n \wedge (N - k)) ] &\text{if}~(t,b) = (n \wedge k, 0), \\
    [k + b, N  - (n \wedge (N-k)) + b] &\text{if}~b \geq 1. 
\end{cases}
\]
Thus, the support is strictly increasing (with respect to the strong set order) in $(t,b)$. 

For $0 \leq t \leq (n \wedge k) -1$, we have
\begin{equation} \label{eq:t_ratio}
		\frac{f(t + 1, 0 | j)}{f(t, 0| j)}  = \frac{h_{n \wedge k}^{k} ( t +1 | j \wedge k)}{h_{n \wedge k}^{k} ( t | j \wedge k)},
\end{equation}
whenever both terms are strictly positive. This ratio is weakly increasing in $j$ by our observation about the hypergeometric distribution. 

For $0 \leq b \leq (n \wedge (N - k))  - 1$, we have 
\begin{equation} \label{eq:b_ratio}
		\frac{f ( n \wedge k, b + 1 | j)}{f( n \wedge k, b | j)} 
		= \frac{h^{N-k}_{n \wedge (N-k)} ( b +1| (j - k)_+) }{h^{N-k}_{n \wedge (N-k)} ( b| (j - k)_+)},
\end{equation}
whenever both terms are strictly positive. This ratio is weakly increasing in $j$ by our observation about the hypergeometric distribution. 

In view of the support comparisons, we conclude that the ratios in \eqref{eq:t_ratio} and \eqref{eq:b_ratio} are weakly increasing in $j$ and nonconstant over the union of the supports of the consecutive events.

\subsection{Proof of Lemma~\ref{res:favorability_thresholds}} \label{sec:proof_fav2}
We use the notation for the hypergeometric distribution introduced in \cref{sec:proof_fav1}.

(i) For $1 \leq n \leq k$ and $0 \leq t \leq n$, let 
	\[
	f (k,n,t | j) = \P ( T_n^k = t \mid \abs{\th} = j ) = h_{n}^{k} (t | j \wedge k).
    \]
    The support of the event $T_n^k = t$ is given by 
    \[
        \begin{cases}
            \{t, \ldots, k - n +t\} &\text{if}~t < n, \\
            \{n, \ldots, N\} &\text{if}~ t = n.
        \end{cases}
    \]
    Thus, this support is strictly decreasing in $n$. If $t < n$, the    support is strictly increasing (with respect to the strong set order) in $k$. 

    First we prove the comparison in $k$. For $n > t$, we have 
    \[
        \frac{f(k+1,n,t| j)}{f(k, n, t | j)} = \frac{ h_n^{k+1} ( t | j)}{h_n^k ( t|  j)} = \frac{ k + 1 - n}{k + 1} \cdot \frac{ k + 1 - j}{k + 1 - j - n + t},
    \]
    whenever both densities are strictly positive.   In view of the support comparisons, we conclude that the ratio is weakly increasing and nonconstant in $j$ over the union of the supports. For $n = t$, we have
       \[
        \frac{f(k+1,n,n| j)}{f(k, n, n | j)} = \frac{ h_n^{k+1} ( n | j \wedge (k+1))}{h_n^k ( n|  j \wedge k)}
        = \begin{cases} 
        \frac{ k + 1 - n}{k +1} &\text{if}~ j \leq k, \\
        1 &\text{if}~ j > k.
        \end{cases}
        \]
  This ratio is weakly increasing (and nonconstant) in $j$ over the support. 
        
    Next, we prove the comparison in $n$. For $n +1 \leq k$, we have
    \[
        \frac{ f(k, n+1, t | j)}{f ( k, n, t | j)} = \frac{ h_{n+1}^{k} ( t| j)}{h_{n}^k ( t | j)} = \frac{ n+1}{k-n} \cdot \frac{ k - j - n + t}{n + 1 - t},
    \]
    whenever both densities are strictly positive. In view of the support comparisons, we conclude that the ratio is weakly decreasing and nonconstant in $j$ over the union of the supports.
    
(ii) Fix $n \in \{ 1, \ldots, N -1\}$. Let
    \begin{equation*}
    \begin{aligned}
      f(k|j ) &= \P \Paren{ (T_n^k, B_n^{N-k})  = (n \wedge k , 0) \mid \abs{\th} = j } \\ 
       &=  h_{n \wedge k}^{k} ( n \wedge k | j \wedge k) h^{N-k}_{n \wedge (N-k)} ( 0| (j - k)_+).
    \end{aligned}
    \end{equation*}
    We have 
    \[
        f(k|j) = 
        \begin{cases}
\frac{\binom{j}{n\wedge k}}
     {\binom{k}{n\wedge k}} &\text{if}~j\leq k, \\[10pt]  
\frac{\binom{N-j}{n\wedge(N-k)}}
     {\binom{N-k}{n\wedge(N-k)}} &\text{if}~j > k.
     \end{cases}
\]
Therefore, over the union of supports, we have
\begin{equation} \label{eq:f_ratio}
\frac{f(k+1 | j)}{f(k | j)}
=
\begin{cases}
a_k &\text{if}~j\leq k,\\
b_k &\text{if}~j\geq k+1,
\end{cases}
\end{equation}
where
\[
a_k=
\begin{cases}
0 &\text{if}~k<n,\\
\frac{k+1-n}{k+1}&\text{if}~k\geq n,
\end{cases}
\]
and
\[
b_k=
\begin{cases}
\frac{N-k}{N-k-n}&\text{if}~k<N-n,\\
\infty &\text{if}~k\geq N-n.
\end{cases}
\]
Since $a_k<1<b_k$, the ratio in \eqref{eq:f_ratio} is weakly increasing and nonconstant
in $j$. Hence the event is strictly increasing in favorability with
$k$.

\subsection{Proof of Theorem~\ref{res:buying_threshold}}

Finally, we consider the attribute threshold for buying. 
If $T_n^k < n \wedge k$, then $B_n^{N-k} = 0$. And by \cref{res:favorability}, we have
\[
    \bar{p}_{k,n} = \E [ v(\th) \mid (T_n^k, B_n^{N-k}) = (n \wedge k,0) ] \leq \E [ v(\th) \mid T_n^k = n \wedge k ].
\]
Therefore, for $p \leq \bar{p}_{k,n}$, we equivalently have
\[
  t^\ast (p,k,n) = \min \Set{ t \in \{ 1, \ldots, n \wedge k \}  : \E [ v(\th) \mid T_n^k = t ] \geq p  }.
\]
Clearly, $t^\ast (p,k,n)$ is weakly increasing in $p$. For $k \geq n$, \cref{res:favorability_thresholds}.\ref{it:t} implies that $\E [ v(\th) \mid T_n^k  = t ]$ is strictly increasing in $k$ and strictly decreasing in $n$, so $t^\ast (p,k,n)$ is weakly decreasing in $k$ and weakly increasing in $n$.


\bibliographystyle{chicago}
\bibliography{literature}

\newpage

\end{document}